\documentclass[12pt, doublecolumn]{IEEEtran}
\usepackage[dvips]{graphicx}
\usepackage{subfigure}
\usepackage{epsfig, psfrag, amsmath, amssymb, amsfonts, latexsym, eucal, balance, hyperref,indentfirst,bookmark}
\newtheorem{thm}{Theorem}
\newtheorem{cor}{Corollary}

\begin{document}
\title{Multi-channel Hybrid Access Femtocells:\\
A Stochastic Geometric Analysis}
\author{Yi Zhong and Wenyi Zhang, \emph{Senior Member, IEEE}
\thanks{The authors are with Department of Electronic Engineering and Information Science, University of Science and Technology of China, Hefei 230027, China (email: geners@mail.ustc.edu.cn, wenyizha@ustc.edu.cn).
The research has been supported by the National Basic
Research Program of China (973 Program) through grant
2012CB316004, National Natural Science Foundation of
China through grant 61071095, Research Fund for the
Doctoral Program of Higher Education of China through
grant 20103402120023, and by MIIT of China through
grants 2010ZX03003-002 and 2011ZX03001-006-01.
}
}
\maketitle
\thispagestyle{empty}
\begin{abstract}
For two-tier networks consisting of macrocells and femtocells, the channel access mechanism can be configured to be open access, closed access, or hybrid access. Hybrid access arises as a compromise between open and closed access mechanisms, in which a fraction of available spectrum resource is shared to nonsubscribers while the remaining reserved for subscribers.
This paper focuses on a hybrid access mechanism for multi-channel femtocells which employ orthogonal spectrum access schemes.
Considering a randomized channel assignment strategy, we analyze the performance in the downlink.
Using stochastic geometry as technical tools, we model the distribution of femtocells as Poisson point process or Neyman-Scott cluster process and derive the distributions of signal-to-interference-plus-noise ratios, and mean achievable rates, of both nonsubscribers and subscribers. The established expressions are amenable to numerical evaluation, and shed key insights into the performance tradeoff between subscribers and nonsubscribers. The analytical results are corroborated by numerical simulations.
\end{abstract}

\setcounter{page}{1}

\section{Introduction}
\label{sec:intro}

In current cellular network services, about 50\% of phone calls and 70\% of data services take place indoors \cite{mansfield2008femtocells}.
For such indoor use cases, network coverage is a critical issue.
One way to improve the indoor performance is to deploy the so-called femtocell access points (FAPs) besides macrocell base stations (MBSs). 
Femtocells are small cellular base stations, typically designed for use in home or small business \cite{chandrasekhar2008femtocell}\cite{claussen2008overview}.
The use of femtocells not only benefits the users, but also the operators.
As the distance between transmitter and receiver is reduced, users will enjoy high quality links and power savings.
Furthermore, the reduced transmission range also creates more spatial reuse and reduces electromagnetic interference.

\begin{figure}
\centering
\includegraphics[width=0.45\textwidth]{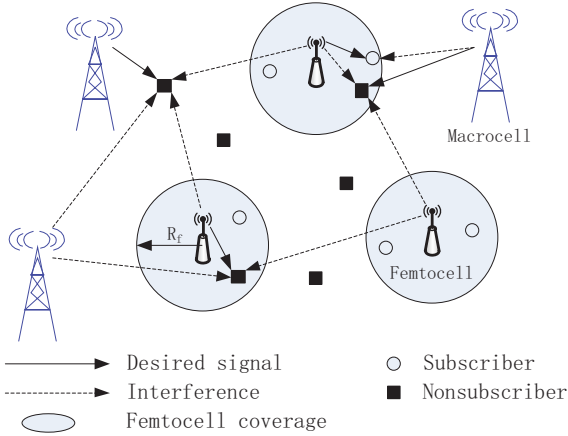}
\caption{Downlink two-tier network model for hybrid access femtocell.}
\label{fig:NETWORK}
\end{figure}

Among the many challenges faced by femtocells, and more generally, two-tier networks, is the issue of interference; see Figure \ref{fig:NETWORK}.
The two-tier interference problem differs from that in traditional single-tier networks in several important aspects:
First, due to the limitations of access mechanism, a user equipment (UE) may not be able to connect to the access point which offers the best service.
Second, since femtocells connect to operator's core network via subscribers' private ISP, coordination between macrocells and femtocells and among femtocells is limited.
Finally, compared to planned macrocell deployments, femtocells are usually deployed in an ad hoc manner, and the randomly placed femtocells make it difficult to manage the interference.
In two-tier networks, interference can be categorized into two types:
(a) cross-tier, referring to the interference from one tier to the other; (b) co-tier, referring to the interference within a tier.

In this paper, we consider two-tier networks based on multicarrier techniques, for example those deploying LTE or WiMAX standards, which use orthogonal frequency-division multiple access (OFDMA) techniques.
In multicarrier systems, the available spectrum is divided into orthogonal subcarriers, which are then grouped into multiple subchannels, assigned to
different users.
Due to the flexibility in channel assignment, the interference may be alleviated.

The access mechanism of femtocells (see, e.g., \cite{nortel4open}) is a key factor that affects the performance of two-tier networks, and generally can be classified as follows, where we call the UEs registered to a femtocell as subscribers, and those not registered to any femtocell as nonsubscribers.
\begin{itemize}
\item \textbf{Closed access:}
An FAP only allows its subscribers to connect.
\item \textbf{Open access:}
An FAP allows all its covered UEs, no matter registered or not, to connect.
\item \textbf{Hybrid access:}
An FAP allows its covered nonsubscribers to connect via a subset of its available subchannels, and reserves the remaining subchannels for its subscribers.
\end{itemize}

Hybrid access \cite{hybrid} is an intermediate approach, in which a fraction of resource is allocated to nonsubscribers.
By doing so, nonsubscribers near an FAP may handover into the femtocell to avoid high interference;
meanwhile, with certain amount of resource reserved for subscribers, the performance of subscribers may be well assured even in the presence of nonsubscribers.

In hybrid access, a central issue is how to allocate the resource between subscribers and nonsubscribers. Previous studies \cite{de2010access} \cite{valcarce2009limited} indicate that hybrid access improves the network performance at the cost of reduced performance for subscribers,
therefore suggesting a tradeoff between the performance of nonsubscribers and subscribers.
In this paper, we consider a hybrid access mechanism that uses a randomized channel assignment strategy, and analyze the performance in the downlink of both macrocells and femtocells.
We employ stochastic geometry to characterize the spatial distributions of users as well as access points; see, e.g., \cite{haenggi2009stochastic} and references therein for its recent applications in wireless networks.
In order to make the work integral, we will carry out the analysis in two different cases.
As a general assumption, we first assume that the FAPs are distributed as a Poisson point process (PPP).
Then, we switch to the case when the FAPs are distributed as a Neyman-Scott cluster process.
The cluster process is likely to be more realistic because the FAPs are typically deployed in populous locations, like commercial or residential area.
Accordingly, we derive the key performance indicators including mean achievable rates and distributions of the signal-to-interference-plus-noise ratios (SINRs) of both nonsubscribers and subscribers.
In our study, we establish general integral expressions for the performance indicators, and closed form expressions under specific model parameters.
With the obtained results, we reveal how the performance of subscribers and nonsubscribers trades off each other.

The introduction of stochastic geometry in the analysis of wireless network is not our original, and an overview of related works is as follows. In \cite{andrews2010tractable},
the authors proposed to study key performance indicators for cellular networks,
such as coverage probabilities and mean achievable rates.
In \cite{chandrasekhar2009spectrum}, the considered scheme divides the spectrum resource into two orthogonal parts which are assigned to macrocells and femtocells, respectively, with femtocells being closed access.
In \cite{chandrasekhar2009uplink}, the authors considered two-tier femtocell networks using time-hopped CDMA, examining the uplink outage probability and the interference avoidance capability.
In \cite{cheung2012throughput}, the success probabilities under Rayleigh fading for both macrocells and femtocells are derived in uplink and downlink respectively.
In \cite{yin2009generalized} and \cite{vaze2011transmission}, stochastic geometry tools are applied in the coexistence analysis of cognitive radio networks.
In \cite{xia2010open} and \cite{jo2010open}, the authors studied the performance of various femtocell access mechanisms, under substantially different system models from ours.
More explicitly, the work in \cite{xia2010open}, which does not make use of stochastic geometry, focused on the uplink with only one MBS and one FAP in the model.
Though the work in \cite{jo2010open} also applies the Laplace transform of interference in the derivation, the substantially difference of our work lies in that we take the load (measured by the number of UEs in a cell) into consideration, utilizing the size distribution of Voronoi cells to derive the distribution of the load. Moreover, we model the mechanism for sharing sub-channels in multi-channel systems, and propose to use a two-scale approximation which substantially simplifies the analysis.
All the works mentioned above are based on the PPP assumption.
The works applying the clustered model can be found in \cite{ganti2009interference} which derived the success probability for transmission in clustered ad-hoc networks and in \cite{gulati2010statistics} which discussed the property of interference with clustered interferers.

The main contribution of our work is detailed as follows.
The existing works mostly ignore the network load which is a key factor that affects the distribution of interfering access points (APs).
For example, when the load is uniformly distributed in the plane, the APs with larger coverage may experience more load, thus leading to more interference to the network.
Moreover, the optimal proportion of shared resources of a femtocell also depends on the distribution of network load.
In our work, we focus on the performance analysis in the context of multi-channel systems, in which case not all sub-channels are occupied and not all APs cause interference to a given subchannel.
We evaluate the two-tier interference when the FAPs are distributed as the PPP and the Neyman-Scott cluster respectively.
In addition, we propose two-scale approximation to simplify the analysis and verify the effectiveness of the approximation by simulation.

The remaining part of the paper is organized as follows.
Section \ref{sect:Network Model} describes the two-tier network model, the channel assignment strategy, and the hybrid access mechanism. Based on a two-scale approximation for the spatial distributions of FAPs, Section \ref{sect:Capacity Analysis} analyzes the statistical behavior of UEs, deriving the distributions of the number of UEs connecting to either an FAP or an MBS, as well as the probabilities of a subchannel being used by either an FAP or an MBS.
Built upon those statistics, Section \ref{sect:AAR} and \ref{sect:Cluster} establish expressions for the distributions of SINRs, and mean achievable rates in the cases when the FAPs are distributed as PPP and Neyman-Scott cluster respectively.
Section \ref{sect:numerical} illustrates the aforementioned analysis by numerical results, which are also corroborated by simulations. Finally, Section \ref{sect:conclusions} concludes the paper.

\section{Network Model}
\label{sect:Network Model}

\subsection{Hybrid Access Femtocells}
\label{subsec:femtocell-description}

\begin{figure}
  \centering
  \subfigure[FAPs are distributed as PPP.]{
    \label{fig:voronoi:a} 
    \includegraphics[width=0.45\textwidth]{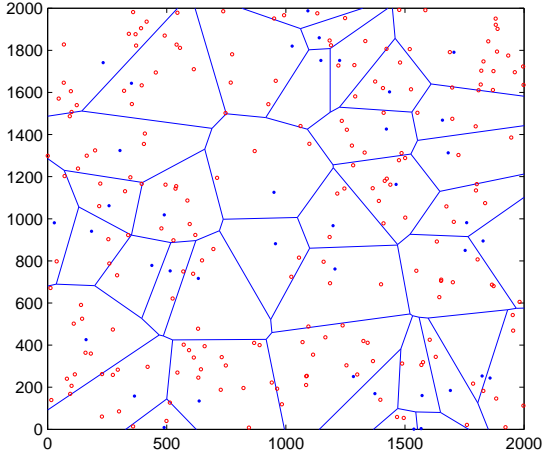}}
  \subfigure[FAPs are distributed as Neyman-Scott cluster.]{
    \label{fig:voronoi:b} 
    \includegraphics[width=0.45\textwidth]{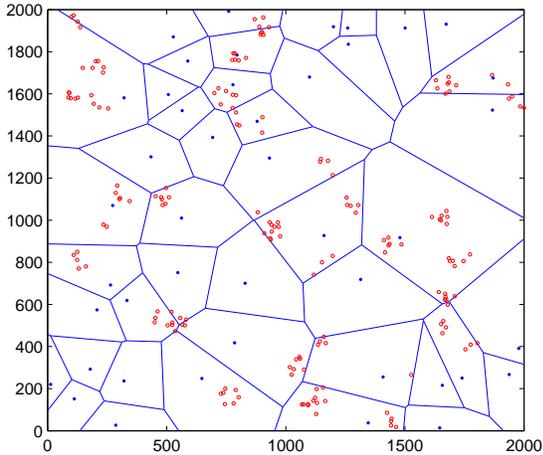}}
  \caption{The Voronoi macrocell topology, in which each Voronoi cell is the coverage area of a macrocell and each small circle represents a femtocell.}
  \label{fig:voronoi} 
\end{figure}

In the two-tier network, we consider two types of access points, MBSs and FAPs. The MBSs constitute the macrocell tier, and they induce a Voronoi tessellation of the plane (see Figure \ref{fig:voronoi}). When a UE attempts to access the macrocell network, it chooses to connect to the MBS in the Voronoi cell in which the UE is situated. An FAP provides network access to UEs in its vicinity, and we assume that all FAPs have a covering radius of $R_f$. Within the covered circular area of each FAP are two types of UEs, called subscribers and inside nonsubscribers. Inside nonsubscribers are those UEs who gather around an FAP without subscribing to its service; for example, transient customers in a shop or a restaurant. Besides those two types of UEs, we also consider a third type of UEs, outside nonsubscribers, who are uniformly scattered over the whole plane, corresponding to those regular macrocell network users.

\begin{figure}
\centering
\includegraphics[width=0.4\textwidth]{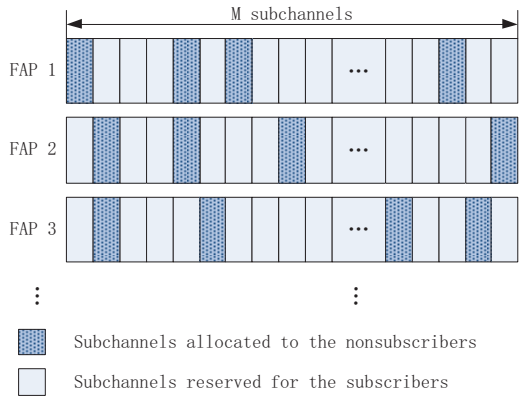}
\caption{Spectrum allocation in each hybrid access femtocell. All the $M_s$ shared subchannels are randomly selected by each FAP.}
\label{fig:SUBCHANNEL}
\end{figure}

The available spectrum is evenly divided into $M$ subchannels, which are to be shared by both macrocell tier and femtocell tier. Each FAP is configured to allocate a fixed number, $M_s$, of subchannels for its covered inside nonsubscribers. These $M_s$ subchannels are called shared subchannels, and the remaining $M_r = M - M_s$ subchannels are called reserved subchannels as they are reserved for the subscribers. In the considered hybrid access mechanism, each FAP selects its shared subchannels randomly, and independently of other FAPs. We assume that each UE, whether subscriber or nonsubscriber, needs one subchannel for transmitting. When a UE accesses an MBS or an FAP, the serving subchannel is selected randomly (see Figure \ref{fig:SUBCHANNEL}).

The hybrid access mechanism operates as follows.
\begin{itemize}
\item
A subscriber accesses to one of the $M_r$ reserved subchannels of its corresponding FAP. When there are more than $M_r$ subscribers in an FAP, they are served by time-sharing with equal time proportion.
\item
An inside nonsubscriber accesses to one of the $M_s$ shared subchannels of its covering FAP. When there are more than $M_s$ inside nonsubscribers in an FAP, they are served by time-sharing with equal time proportion.
\item
An outside nonsubscriber accesses the MBS located in the Voronoi cell in which the outside nonsubscriber is situated. When there are more than $M$ outside nonsubscribers in the Voronoi cell, they are served by time-sharing with equal proportion.
\end{itemize}

\subsection{Mathematical Model and Two-scale Approximation}
\label{subsec:two-scale}

To formulate the aforementioned hybrid access scenario mathematically, we model the spatial distributions of the nodes using spatial point processes as follows. The MBSs constitute a homogeneous Poisson point process (PPP) $\Phi_m$ of intensity $\lambda_m$ on the plane.
The distribution of FAPs will be divided into two cases for discussion:
\begin{itemize}
\item
Case 1: the FAPs constitute another homogeneous PPP $\Phi_{f}$ of intensity $\lambda_f$.
\item
Case 2: the FAPs are distributed as a Neyman-Scott cluster process $\Phi_{f}$ \cite{stoyan1995stochastic}.
The center of the clusters are assumed to be distributed according to a stationary PPP $\Phi_{p}$ of intensity $\lambda_p$, which is called the parent process.
For each cluster center $x\in\Phi_{p}$, the FAPs are distributed according to an independent PPP $\Phi^{x}$ of intensity $\lambda_c$ in the circular covered area of radius $R_c$ around the center $x$.
The complete distribution of all FAPs is given as
\begin{equation}
\Phi_{f}=\bigcup_{x\in\Phi_{p}}\Phi^{x}. \label{equ:FAPs}
\end{equation}
In this case, the number of FAPs in a typical cluster is a Poisson random with parameter  $\pi R_c^2\lambda_c$ and the intensity of all FAPs is $\lambda_f=\pi R_c^2\lambda_c\lambda_p$.
\end{itemize}

In the circular covered area of radius $R_f$ of each FAP, the subscribers are distributed according to a homogeneous PPP of intensity $\lambda_{s}$, and the inside nonsubscribers are distributed according to another homogeneous PPP of intensity $\lambda_\mathrm{in}$. Outside nonsubscribers constitute on the whole plane a homogeneous PPP of intensity $\lambda_\mathrm{out}$. All the PPPs are mutually independent.

In this paper, we focus on the downlink performance. The transmit power is set to a constant value $P_m$ for an MBS, and $P_f$ for an FAP. For the sake of convenience, we adopt a standard path loss propagation model with path loss exponent $\alpha > 2$.
Regarding fading, we assume that the link between the serving access point (either an MBS or an FAP) and the served UE experiences Rayleigh fading with parameter $\mu$.
The received signal power of a UE at a distance $r$ from its serving access point therefore is $P_mhr^{-\alpha}$ (MBS) or $P_fhr^{-\alpha}$ (FAP) where $h\sim\mathrm{Exp}(\mu)$.
The fading of interference links may follow an arbitrary probability distribution, and is denoted by $g$.
Furthermore, considering the typical scenario of indoor femtocell deployment, we introduce a wall isolation at the boundary of each FAP coverage area, which incurs a wall penetration loss factor $W < 1$. For all receivers, the noise power is $\sigma^2$.

The different point processes corresponding to different entities in the network interact in a complicated way, thus making a rigorous statistical analysis extremely difficult. For example, an inside nonsubscriber may be covered by more than one FAPs, thus leading to the delicate issue of FAP selection, and furthermore rendering the subchannel usage distributions among FAPs and MBSs intrinsically correlated. To overcome the technical difficulties due to spatial interactions, in the subsequent analysis we propose a two-scale approximation for the network model, motivated by the fact that the covered area of an FAP is significantly smaller than that of an MBS. The two-scale approximation consists of two views, the macro-scale view and the micro-scale view. The macro-scale view concerns an observer outside the coverage area of an FAP, and in that view the whole coverage area of the FAP shrinks to a single point, marked by the numbers of subscribers and inside nonsubscribers therein. The micro-scale view concerns an observer inside the coverage area of an FAP, and in that view the coverage area is still circular with radius $R_f$ in which the subscribers and inside nonsubscribers are spatially distributed. By such a two-scale approximation, an inside nonsubscriber can only be covered by a unique FAP, and the coverage area of an FAP can only be within a unique Voronoi cell of an MBS.
Meanwhile, at the cell edge the outside nonsubscribers are clearly divided by the boundary and the subscribers and inside nonsubscribers are all attached to the corresponding FAPs which are also clearly divided.
These consequences substantially simplify the performance analysis.
In Section \ref{sect:numerical}, we validate the two-scale approximation method through comparing analytical results and simulation results, for network parameters of practical interest.

\section{Statistics of UEs and Subchannels}
\label{sect:Capacity Analysis}

In this section, we characterize the distributions of UEs connecting to different types of access points, and the distributions of used subchannels in MBSs and FAPs. The analysis is based on a snapshot of the network model, and the obtained results will then be applied for characterizing the distributions of SINRs and achievable rates in Section \ref{sect:AAR}.

\subsection{Distributions of UEs}

Let $U_s$ be the number of subscribers accessing a given FAP and from our model we have $U_s\sim\mathrm{Poisson}(\lambda_s\pi R_f^2)$. Similarly, let $U_\mathrm{in}$ be the number of inside nonsubscribers accessing a given FAP, and we have $U_\mathrm{in}\sim\mathrm{Poisson}(\lambda_\mathrm{in}\pi R_f^2)$.

The number of outside nonsubscribers who access a given MBS, denoted by $U_\mathrm{out}$, is characterized as follows. We note that the macrocell coverage area is a Voronoi cell, and denote by $S$ the area of the Voronoi cell. There is no known closed form expression of the probability density function (pdf) of $S$, whereas a simple approximation \cite{ferenc2007size} has proven sufficiently accurate for practical purposes.
Considering scaling, the approximate pdf of the size of a macrocell coverage area is given by
\begin{equation}
f(S)=\frac{343}{15}\sqrt{\frac{7}{2\pi}}(S\lambda_m)^{\frac{5}{2}}\exp(-\frac{7}{2}S\lambda_m)\lambda_m. \label{equ:fs}
\end{equation}

Conditioning upon $S$, the number of outside nonsubscribers is a Poisson random variable with mean $\lambda_\mathrm{out} S$. The probability generating function of the unconditioned $U_\mathrm{out}$ is thus given by
\begin{equation}
G(z)=\int_0^\infty\exp\Big(\lambda_\mathrm{out}(z-1)S\Big)f(S)dS.
\end{equation}
Plugging in the approximate pdf of $S$ and simplifying the integral, we get
\begin{equation}
G(z)=\frac{343}{8}\sqrt{\frac{7}{2}}\Big(\frac{7}{2}-\frac{\lambda_\mathrm{out}}{\lambda_m}(z-1)\Big)^{-\frac{7}{2}}.
\end{equation}
The distribution of $U_\mathrm{out}$ is therefore given by the derivatives of $G(z)$,
\begin{equation}
\mathbb{P}\{U_\mathrm{out}=i\}=\frac{G^{(i)}(0)}{i!}, \quad i = 0, 1, \ldots. \label{equ:U_n_m}
\end{equation}

\subsection{Distributions of Subchannel Usage}

Since the subchannels are uniformly and independently selected by each FAP, it suffices to analyze an arbitrary one of them.
Let us examine the probability that a given subchannel is used by an MBS or an FAP.
First, we evaluate the average number of subchannels used by an MBS or an FAP, and then
we normalize the average number by the total number of subchannels, $M$.

The probability that a subchannel is used by an FAP is
\begin{eqnarray}
P_{\mathrm{busy},f}=\frac{1}{M}\Big(\sum_{i=0}^\infty\min\{i,M_r\}\mathbb{P}\{U_s=i\}\nonumber\\
+\sum_{j=0}^\infty\min\{j,M_s\}\mathbb{P}\{U_\mathrm{in}=j\}\Big).
\end{eqnarray}

For a Poisson random variable $N\sim\mathrm{Poisson}(\lambda)$, its cumulative distribution function (cdf) is
\begin{equation}
\sum_{i=0}^n\mathbb{P}\{N=i\}=\sum_{i=0}^n\frac{\lambda^i}{i!}e^{-\lambda}=\frac{\Gamma(n+1,\lambda)}{n!}, \label{equ:poissoncdf}
\end{equation}
where $\Gamma(s,x)=\int_x^\infty t^{s-1}e^{-t}\mathrm{d}t$ is the incomplete gamma function. Using (\ref{equ:poissoncdf}) to simplify $P_{\mathrm{busy},f}$, we get
\begin{eqnarray}
P_{\mathrm{busy},f}=1+\frac{M_r}{M}\frac{1}{M_r!}\Big(\lambda_s\pi R_f^2\Gamma(M_r,\lambda_s\pi R_f^2)\nonumber\\
-\Gamma(M_r+1,\lambda_s\pi R_f^2)\Big) \nonumber\\
+\frac{M_s}{M}\frac{1}{M_s!}\Big(\lambda_\mathrm{in}\pi R_f^2\Gamma(M_s,\lambda_\mathrm{in}\pi R_f^2)\nonumber\\
-\Gamma(M_s+1,\lambda_\mathrm{in}\pi R_f^2)\Big).
\end{eqnarray}

The probability that a subchannel is used by an MBS is
\begin{eqnarray}
P_{\mathrm{busy},m}=\frac{1}{M}\sum_{i=0}^\infty\min\{i,M\}\mathbb{P}\{U_\mathrm{out}=i\},
\end{eqnarray}
where $\mathbb{P}\{U_\mathrm{out}=i\}$ is given by (\ref{equ:U_n_m}).

The spatial point process of FAPs that use a given subchannel is the independent thinning of the original process of FAPs $\Phi_{f}$ by the probability $P_{\mathrm{busy}, f}$, denoted by $\Phi_{f}'$.
The term ``independent thinning'' means that $\Phi_{f}'$ can be viewed as obtained from $\Phi_{f}$ by independently removing points with probability $1 - P_{\mathrm{busy},f}$.
When the FAPs are distributed as a homogeneous PPP, the resulting point process is still a homogeneous PPP with intensity $\lambda_f'=\lambda_f P_{\mathrm{busy},f}$.
For the case when the FAPs are distributed as Neyman-Scott cluster process,
the resulting point process is a Neyman-Scott cluster process with intensity $\lambda_f'=\lambda_f P_{\mathrm{busy},f}$.
Moreover, the intensity of the parent process is still $\lambda_p$ and the intensity of the FAPs in each cluster is reduced to $\lambda_c'=\lambda_cP_{\mathrm{busy},f}$.
As for the MBSs, the correlations between the sizes of neighboring cells may lead to the dependence in the thinning of the original PPP of MBSs.
However, in order to facilitate the analysis, we assume that the independent thinning assumption still holds.
Therefore, the spatial process of MBSs that use a given subchannel is the independent thinning of the original PPP of MBSs $\Phi_{m}$ by the probability $P_{\mathrm{busy}, m}$, denoted by $\Phi_{m}'$ with intensity $\lambda_m'=\lambda_m P_{\mathrm{busy},m}$.
Meanwhile, the computation of $P_{\mathrm{busy}, m}$ and the SINR distribution can also be decoupled.
The simulation results validate that our independent thinning assumption is rather reliable.
These two independently thinned point processes will prove useful in the subsequent analysis.

\section{Performance With Poisson Distribution Of FAPs}
\label{sect:AAR}

In this section, we derive the distributions of the SINRs and the mean achievable rates of UEs served by MBS and FAP respectively when the FAPs are distributed as the PPP.
For each type of UEs, we begin with general settings, and then simplify the general results under specific parameters to gain insights.
The mean achievable rates are the averaged instantaneous achievable rates over both channel fading and spatial distributions of UEs and access points.

\subsection{Macrocell UEs}
\subsubsection{General Case}
For an active UE served by an MBS, it must be occupying one subchannel of the MBS.
The following theorem gives the cdf of SINR and the mean achievable rate of each active macrocell UE in general case.
\begin{thm}
\label{thm:sinr macrocell}
The cdf of the SINR of a macrocell UE, denoted by $Z_m(T)=\mathbb{P}\{\mathrm{SINR}\leq T\}$, is given by
\begin{eqnarray}
Z_m(T)=1-\pi \lambda_m\!\!\!\int_0^\infty\!\!\!\exp\Big(\pi v\lambda_m'\Big(1-\beta(T,\alpha)\nonumber\\
-\frac{1}{P_{\mathrm{busy}, m}}\Big)-\frac{\mu Tv^{\frac{\alpha}{2}}\sigma^2}{P_m}\Big)\mathrm{d}v. \label{equ:OUT_SINR_DIS}
\end{eqnarray}
The mean achievable rate of a macrocell UE is given by
\begin{eqnarray}
\tau_m=\pi\lambda_m\!\!\!\int_0^\infty\!\!\!\int_0^\infty\!\!\! \exp\Big(\pi v\lambda_m'\Big(1-\beta(e^t-1,\alpha)\nonumber\\
-\frac{1}{P_{\mathrm{busy}, m}}\Big)-\frac{\mu v^\frac{\alpha}{2}\sigma^2(e^t-1)}{P_m}\Big)\mathrm{d}t\mathrm{d}v.\label{equ:OUT_RATE}
\end{eqnarray}
In (\ref{equ:OUT_SINR_DIS}) and (\ref{equ:OUT_RATE}), $\beta(T, \alpha)$ is given by
\begin{eqnarray}
\beta(T,\alpha)=\frac{2(\mu T)^{\frac{2}{\alpha}}}{\alpha}\mathbb{E}_g\Big(g^\frac{2}{\alpha}\Big(\Gamma(-\frac{2}{\alpha},\mu Tg)\nonumber\\
-(1+\frac{\lambda_f'(WP_f)^{\frac{2}{\alpha}}}{\lambda_m'P_m^{\frac{2}{\alpha}}})\Gamma(-\frac{2}{\alpha})\Big)\Big). \label{equ:beta}
\end{eqnarray}
\end{thm}

The proof of Theorem \ref{thm:sinr macrocell} is in Appendix \ref{appendix:a}.

In Theorem \ref{thm:sinr macrocell}, $Z_m(T)$ in (\ref{equ:OUT_SINR_DIS}) gives the probability that the SINR is below a given target level $T$, and $\tau_m$ in (\ref{equ:OUT_RATE}) gives the mean achievable rate of a macrocell UE. The integrals in (\ref{equ:OUT_SINR_DIS}) and (\ref{equ:OUT_RATE}) can be evaluated by numerical methods, and furthermore, they can be simplified to concise forms in the special case when the interference experiencing Rayleigh fading and the path loss exponent being $\alpha=4$ with no noise, $\sigma^2=0$.

\subsubsection{Special Case When Interference Experiences Rayleigh Fading}

Here we consider the case where the interference experiences Rayleigh distribution with mean $\mu$, i.e. $g\sim \mathrm{Exp}(\mu)$.
In this case, the results are as follows.
\begin{cor}
\label{cor:sinr macrocell sim}
When the interference follows Rayleigh fading, the cdf of the SINR is
\begin{eqnarray}
\!\!\!Z_m(T)&\!\!\!\!\!\!=\!\!\!\!\!\!&1-\pi \lambda_m\!\!\!\int_0^\infty\!\!\!\exp\Big(-\pi v\lambda_m-\pi v\lambda_m'\varphi(T,\alpha) \nonumber \\
\!\!\!&&-\pi v\lambda_f'\Big(\frac{P_fWT}{P_m}\Big)^{2/\alpha}\Gamma(1+\frac{2}{\alpha})\Gamma(1-\frac{2}{\alpha})\nonumber\\
\!\!\!&&-\frac{\mu Tv^{\alpha/2}\sigma^2}{P_m}\Big)\mathrm{d}v.
\end{eqnarray}
The mean achievable rate is
\begin{eqnarray}
\!\!\!\tau_m&\!\!\!\!\!\!=\!\!\!\!\!\!&\pi \lambda_m\int_0^\infty\!\!\!\int_0^\infty\!\!\!\exp\Big(-\pi v\lambda_m-\pi v\lambda_m'\varphi(e^t-1,\alpha) \nonumber \\
\!\!\!&&-\pi v\lambda_f'\Big(\frac{P_fW(e^t-1)}{P_m}\Big)^{2/\alpha}\Gamma(1+\frac{2}{\alpha})\Gamma(1-\frac{2}{\alpha})\nonumber\\
\!\!\!&&-\frac{\mu v^{\alpha/2}\sigma^2(e^t-1)}{P_m}\Big)\mathrm{d}v\mathrm{d}t,
\end{eqnarray}
where
\begin{equation}
\varphi(T,\alpha)=T^{2/\alpha}\int_{T^{-2/\alpha}}^\infty\frac{1}{1+u^{\alpha/2}}\mathrm{d}u. \label{equ:varphi}
\end{equation}
\end{cor}

The results in Corollary \ref{cor:sinr macrocell sim} are further simplified in the following special cases.

Specifically, when $\alpha=4$ we obtain
\begin{eqnarray}
Z_m(T)&\!\!\!\!\!\!=\!\!\!\!\!\!&\nonumber\\
&&\!\!\!\!\!\!\!\!\!\!\!\!\!\!\!\!\!\!\!\!\!\!\!\!\!\!\!\!1-\frac{1}{1+\sqrt{T}\Big(\arctan\sqrt{T}+\frac{\pi}{2}\frac{\lambda_f'} {\lambda_m'}\sqrt{\frac{WP_f}{P_m}}\Big)P_{\mathrm{busy},m}}.\nonumber\\
&& \label{equ:SINR_Simple}
\end{eqnarray}
The mean achievable rate of a macrocell UE is simplified into
\begin{equation}
\tau_m=\int_0^{\frac{\pi}{2}}\frac{2}{\tan y +(\frac{\pi}{2}-y)P_{\mathrm{busy},m}+\frac{\pi}{2}\frac{\lambda_f'}{\lambda_m}\sqrt{\frac{WP_f}{P_m}}}\mathrm{d}y.
\end{equation}

From a practical perspective, it is desirable to shape $Z_m(T)$ to make it small for small values of $T$. From (\ref{equ:SINR_Simple}), we see that there are two approaches to shape $Z_m(T)$. First, $Z_m(T)$ decreases as $P_{\mathrm{busy},m}$, the probability that a subchannel is used by an MBS, decreases. This may be interpreted as an effect of frequency reuse. Second, $Z_m(T)$ decreases as the whole term, $\frac{\lambda_f'}{\lambda_m}\sqrt{\frac{WP_f}{P_m}}$, decreases, which corresponds to a number of network parameters, representing the effect due to the deployment of the femtocell tier.

\subsection{Femtocell UEs}
\subsubsection{General Case}

A UE served by an FAP occupies one subchannel of the FAP.
The following theorem gives the cdf of the SINR and the mean achievable rate of each active femtocell UE in general case.
\begin{thm}
\label{thm:sinr femtocell}
The cdf of the SINR of a femtocell UE in general case is given by
\begin{eqnarray}
Z_f(T) = 1-\frac{1}{R_f^2}\int_0^{R_f^2}\exp\Big(-\rho(\alpha)T^{2/\alpha}v\nonumber\\
-\frac{\mu Tv^{\alpha/2}\sigma^2}{P_f}\Big)\mathrm{d}v. \label{equ:SINR_N_F}
\end{eqnarray}
The mean achievable rate of a femtocell UE is given by
\begin{eqnarray}
\label{eqn:tauf}
\tau_f = \frac{1}{R_f^2}\int_0^{R_f^2}\!\!\!\int_0^\infty\!\! \exp\Big(-\rho(\alpha)(e^t-1)^{2/\alpha}v\nonumber\\
-\frac{\mu v^{\alpha/2}\sigma^2(e^t-1)}{P_f}\Big)\mathrm{d}t\mathrm{d}v.
\end{eqnarray}
in (\ref{equ:SINR_N_F}) and (\ref{eqn:tauf}), $\rho(\alpha)$ is given by
\begin{eqnarray}
\rho(\alpha)=-\frac{2\pi\mu ^{2/\alpha}}{\alpha}\Gamma\Big(-\frac{2}{\alpha}\Big)\Big(\lambda_m'\Big(\frac{WP_m}{P_f}\Big)^{2/\alpha}\nonumber\\
+\lambda_f'W^{4/\alpha}\Big)\mathbb{E}_g(g^{2/\alpha}). \label{equ:rho}
\end{eqnarray}
\end{thm}

The proof of Theorem \ref{thm:sinr femtocell} is in Appendix \ref{appendix:c}.

\subsubsection{Special Case When Interference Experiences Rayleigh Fading}

For Rayleigh fading, the results are essentially the same as that in the general fading case.
The only additional simplification is the evaluation of $\rho(\alpha)$.
We just give the result in the very special case when $\sigma^2=0$ and $\alpha=4$.

When the interference experiences Rayleigh fading, we have $\mathbb{E}_g(g^\frac{1}{2})=\mu\int_0^\infty g^\frac{1}{2}e^{-\mu g}\mathrm{d}g=\frac{1}{2}\sqrt{\frac{\pi}{\mu}}$, and consequently,
\begin{equation}
\rho(4)=\frac{\pi^2}{2}\Big(\lambda_m'\sqrt{\frac{WP_m}{P_f}}+\lambda_f'W\Big).
\end{equation}
The SINR distribution is then given by
\begin{equation}
Z_f(T)=1-\frac{1-e^{-\rho(4)\sqrt{T}R_f^2}}{\rho(4)\sqrt{T}R_f^2}.
\end{equation}
Let $y=\rho(4)R_f^2\sqrt{e^t-1}$, the mean achievable rate is simplified into
\begin{equation}
\tau_f=2\int_0^\infty\frac{1-e^{-y}}{y^2+\rho^2(4) R_f^4}\mathrm{d}y.
\end{equation}
These expressions are convenient for numerical evaluation.

\section{Performance With Clustered FAPs}
\label{sect:Cluster}
In this section, we derive the distributions of SINRs and the mean achievable rates in the case when the FAPs are distributed as the Neyman-Scott cluster process.
To facilitate the analysis, we only focus on the case when the interference links experience exponential fading.

\subsection{Macrocell UEs}
The following theorem gives the cdf of SINR and the mean achievable rate of each active macrocell UE.
\begin{thm}
\label{thm:sinr macrocell_cluster}
In the case when the FAPs are distributed as a Neyman-Scott cluster process and the interference experiences Rayleigh fading, the cdf of the SINR of a macrocell UE, denoted by $Z_m(T)=\mathbb{P}\{\mathrm{SINR}\leq T\}$, is given by
\begin{eqnarray}
Z_m(T)=1-\pi\lambda_m\!\!\!\int_0^\infty\!\!\!\exp\bigg(-\pi v\lambda_m\nonumber\\
-\lambda_p\int_{R^2}\Big(1-\eta\Big(\frac{Tv^{\alpha/2}WP_f}{P_m},x\Big)\Big)\mathrm{d}x \nonumber \\
-\frac{\mu Tv^{\alpha/2}\sigma^2}{P_m}-\pi v\lambda_m'\varphi(T,\alpha)\bigg)\mathrm{d}v. \label{equ:OUT_SINR_DIS_CLUSTER}
\end{eqnarray}
The mean achievable rate of a macrocell UE is
\begin{eqnarray}
\tau_m=\pi \lambda_m\int_0^\infty\!\!\!\int_0^\infty\!\!\!\exp\bigg(-\pi v\lambda_m\nonumber\\
-\lambda_p\int_{R^2}\Big(1-\eta\Big(\frac{ (e^t-1)v^{\alpha/2}WP_f}{P_m},x\Big)\Big)\mathrm{d}x \nonumber \\
-\frac{\mu (e^t-1)v^{\alpha/2}\sigma^2}{P_m}-\pi v\lambda_m'\varphi(e^t-1,\alpha)\bigg)\mathrm{d}v\mathrm{d}t,\label{equ:OUT_RATE_DIS_CLUSTER}
\end{eqnarray}
where $\varphi(T,\alpha)$ is given by (\ref{equ:varphi}). Let $C(o,R_c)$ be the circle centered at the origin with radius $R_c$ and $\eta(s,x)$ is given as
\begin{equation}
\eta(s,x)=\exp\Big(\int_{C(o,R_c)}\frac{-\lambda_c'}{1+\frac{1}{s}|x+y|^\alpha}\mathrm{d}y\Big). \label{equ:eta}
\end{equation}
\end{thm}

The proof of Theorem \ref{thm:sinr macrocell_cluster} is in Appendix \ref{appendix:e}.

\subsection{Femtocell UEs}
The following theorem gives the cdf of SINR and the mean achievable rate of each active femtocell UE.
\begin{thm}
\label{thm:sinr femtocell_cluster}
In the case when the FAPs are distributed as a Neyman-Scott cluster process and the interference experiences Rayleigh fading, the cdf of the SINR of a femtocell UE, denoted by $Z_f(T)=\mathbb{P}\{\mathrm{SINR}\leq T\}$, is given by
\begin{eqnarray}
Z_f(T)=1-\frac{2}{\pi R_c^2R_f^2}\!\!\int_0^{R_f}\!\!\exp\bigg(-\frac{\mu Tr^{\alpha}\sigma^2}{P_f}\nonumber\\
-\pi r^2\lambda_m'\Big(\frac{WTP_m}{P_f}\Big)^{2/\alpha}\Gamma(1+\frac{2}{\alpha})\Gamma(1-\frac{2}{\alpha}) \nonumber \\
-\lambda_p\int_{R^2}\Big(1-\eta(Tr^{\alpha}W^2,x)\Big)\mathrm{d}x\bigg)\nonumber\\
\bigg(\int_{C(o,R_c)}\eta(
Tr^{\alpha}W^2,y-z)\mathrm{d}y\bigg)r\mathrm{d}r. \label{equ:IN_SINR_DIS_CLUSTER}
\end{eqnarray}
The mean achievable rate of a femtocell UE is given by
\begin{eqnarray}
\tau_f=\frac{2}{\pi R_c^2R_f^2}\!\!\int_0^\infty\!\!\int_0^{R_f}\!\!\exp\bigg(-\frac{\mu (e^t-1)r^{\alpha}\sigma^2}{P_f}-\nonumber\\
\pi r^2\lambda_m'\Big(\frac{W(e^t-1)P_m}{P_f}\Big)^{2/\alpha}\Gamma(1+\frac{2}{\alpha})\Gamma(1-\frac{2}{\alpha}) \nonumber \\
-\lambda_p\int_{R^2}\Big(1-\eta\big((e^t-1)r^{\alpha}W^2,x\big)\Big)\mathrm{d}x\bigg)\nonumber\\
\bigg(\int_{C(o,R_c)}\eta\Big( (e^t-1)r^{\alpha}W^2,y-z\Big)\mathrm{d}y\bigg)r\mathrm{d}r\mathrm{d}t, \label{equ:IN_RATE_DIS_CLUSTER}
\end{eqnarray}
where $z=(r,0)$ and $\eta(s,x)$ is given by (\ref{equ:eta}).
\end{thm}
The proof of Theorem \ref{thm:sinr femtocell_cluster} is in Appendix \ref{appendix:f}.

\section{Mean Achievable Rates of Nonsubscribers and Subscribers}

There are two types of nonsubscribers, outside nonsubscribers who access MBSs and inside nonsubscribers who access FAPs.
When the number of outside nonsubscribers in a macrocell is no greater than the total number of subchannels (i.e., $U_\mathrm{out}\leq M$), each nonsubscriber UE exclusively occupies a subchannel, and its mean achievable rate is $\tau_m$.
However, when $U_\mathrm{out}>M$, those $U_\mathrm{out}$ UEs share the $M$ subchannels with mean achievable rate $\frac{M}{U_\mathrm{out}}\tau_m$.
Since the evaluation is conditioned upon the existence of at least one UE, the mean achievable rate of an outside nonsubscriber UE is given by
\begin{equation}
\!\!\!\!\!\!\tau_\mathrm{out}=\frac{\sum_{i=1}^M\mathbb{P}\{U_\mathrm{out}=i\}+\sum_{i=M+1}^\infty\mathbb{P}\{U_\mathrm{out}=i\}\frac{M}{i}}{1-\mathbb{P}\{U_\mathrm{out}=0\}}\tau_m.
\end{equation}
Similarly, the mean achievable rate of an inside nonsubscriber is given by
\begin{equation}
\!\!\!\!\!\!\tau_\mathrm{in}=\frac{\sum_{j=1}^{M_s}\mathbb{P}\{U_\mathrm{in}=j\}+\sum_{j=M_s+1}^\infty\mathbb{P}\{U_\mathrm{in}=j\}\frac{M_s}{j}}{1-\mathbb{P}\{U_\mathrm{in}=0\}}\tau_f.
\end{equation}
When averaged over both inside and outside nonsubscribers, the overall mean achievable rate of nonsubscriber is obtained as
\begin{equation}
\!\!\!\!\tau_n=\frac{\lambda_{out}\tau_\mathrm{out}+\lambda_f\lambda_{in}\pi R_f^2\tau_\mathrm{in}}{\lambda_{out}+\lambda_f\lambda_{in}\pi R_f^2}.\label{equ:tau n}
\end{equation}

Regarding subscribers, note that they are exclusively served by FAPs.
Similar to the analysis of nonsubscribers, the mean achievable rate of a subscriber is given by
\begin{equation}
\tau_s=\frac{\sum_{i=1}^{M_r}\mathbb{P}\{U_s=i\}+\sum_{i=M_r+1}^\infty\mathbb{P}\{U_s=i\}\frac{M_r}{i}}{1-\mathbb{P}\{U_s=0\}}\tau_f. \label{equ:tau s}
\end{equation}

\section{Numerical Results}
\label{sect:numerical}

The numerical results are obtained according to both the analytical results we have derived and Monte Carlo simulation.
The default configurations of system model are as follows (also see Table \ref{table:parameter}).
The total number of subchannels is $M=20$ and the coverage radius of each femtocell is $R_f=10$m.
The transmit power of FAP is $P_f=13$dBm, and that of MBS is $P_m=39$dBm.
We set the path loss exponent as $\alpha=4$, with all links experiencing Rayleigh fading of normalized $\mu = 1$. The wall penetration loss is set as $W = -6$dB. We focus on the interference-limited regime, and for simplicity we ignore the noise power (i.e., $\sigma^2 =0$).
The intensity of MBSs is set as $\lambda_m=0.00001$ and of FAPs $\lambda_f=0.0001$. In the clustered case, the intensity of parent process is set as $\lambda_p=0.00001$ and of FAPs in each cluster $\lambda_c=0.00127$. By setting the radius of each cluster as $R_c=50$m, we get the intensity of FAPs as $\lambda_f=0.0001$. So the average coverage area of an MBS is roughly equal to a circle with a radius of $180$m, and on average there are ten FAPs within the coverage area of an MBS. Unless otherwise specified, the subscribers and inside nonsubscribers are distributed within an FAP coverage are with intensities $\lambda_s=\lambda_\mathrm{in}=0.015$.
The intensity of outside nonsubscribers is set as $\lambda_\mathrm{out}=0.0001$.

\begin{table*}
\centering
\caption{SYSTEM PARAMETERS}
\label{table:parameter}
\begin{tabular}{c|c|c}
\hline
\textbf{Symbol} & \textbf{Description} & \textbf{Value}\\
\hline
$\Phi_m,\Phi_f$ & Point processes defining the MBSs and FAPs & N/A\\
$\lambda_m$ & Density of MBSs & 0.00001 MBS/m$^2$\\
$\lambda_f$ & Density of FAPs & 0.0001 FAP/m$^2$\\
$\lambda_p$ & Density of parent process for the clustered FAPs & 0.00001 center/m$^2$\\
$\lambda_c$ & Density of FAPs in each cluster & 0.00127 FAPs/m$^2$\\
$R_c$ & Radius of each cluster & $50$m\\
$R_f$ & Radius of femtocell & $10$m\\
$\lambda_\mathrm{out}$ & Densities of outside nonsubscribers & 0.0001 user/m$^2$\\
$\lambda_s,\lambda_\mathrm{in}$ & Densities of subscribers and inside nonsubscribers & 0.015 user/m$^2$\\
$P_m,P_f$ & Transmit power at MBS and FAP & $39$dBm,$13$dBm\\
$M$ & Number of subchannels at each access point & $20$\\
$M_s,M_r$ & Number of subchannels shared and reserved by each femtocell & Not fixed\\
$\alpha$ & Path loss exponent & $4$\\
$\mu$ & Rayleigh fading parameter & $1$ (normalized)\\
$W$ & Wall penetration loss & $-6$dB\\
$\sigma^2$ & Noise power & $0$ (interference-limited regime)\\
$P_\mathrm{busy,m},P_\mathrm{busy,f}$ & Probabilities that a given subchannel is used by an MBS and an FAP & Not fixed\\
$\Phi_m',\Phi_f'$ & Point processes defining MBSs and FAPs that interfere a given subchannel & N/A\\
$\lambda_m',\lambda_f'$ & Densities of MBSs and FAPs that interfere a given subchannel & Not fixed\\
$U_\mathrm{out},U_\mathrm{in}$ & Numbers of nonsubscribers that access a given MBS and a given FAP & Not fixed\\
$U_s$ & Numbers of subscribers that access a given FAP & Not fixed\\
$\tau_f,\tau_m$ & Mean achievable rates of femtocell UEs and macrocell UEs & Not fixed\\
$\tau_s,\tau_n$ & Mean achievable rates of subscribers and nonsubscribers & Not fixed\\
\hline
\end{tabular}
\end{table*}

\begin{figure}
\centering
\includegraphics[width=0.45\textwidth]{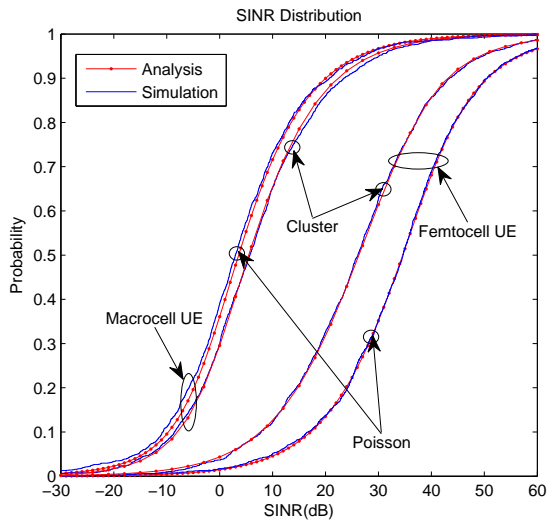}
\caption{Cdfs of SINRs for macrocell UEs and femtocell UEs, when $M_s=10$.}
\label{fig:SINR_cdf}
\end{figure}

Figure \ref{fig:SINR_cdf} displays the SINR distributions of macrocell UEs and femtocell UEs, when the number of shared subchannels is set as $M_s=10$. We plot in dotted curves the analytical results, and we also plot the empirical cdfs obtained from Monte Carlo simulation. The curves reveal that the simulation results match the analytical results well, thus corroborating the accuracy of our theoretical analysis.
From the SINR distributions, we observe that while the macrocell UEs experience a fair amount of interference, due to the shrinking cell size, the interference for femtocell UEs is substantially alleviated.
We also observe that the performance of femtocell UEs is worse in the clustered case than in the Poisson case when the intensity of the FAPs is set as the same value; however, the performance of the macrocell UEs is just the reverse.
This result reveals that the gathering of FAPs leads to more interference to femtocell UEs and at the same time reduces the chance that a macrocell UE being interfered by the nearby FAPs.

\begin{figure}
\centering
\includegraphics[width=0.45\textwidth]{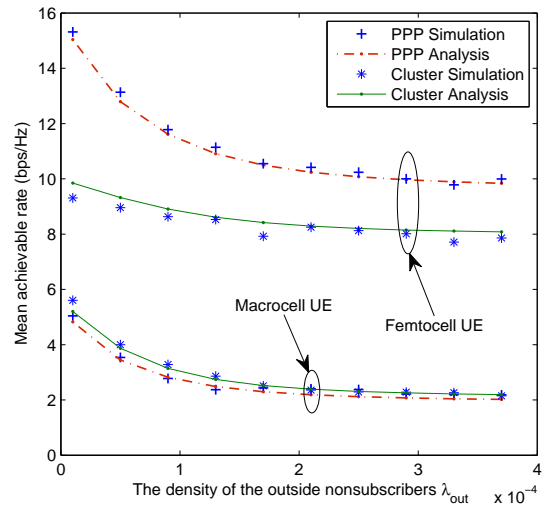}
\caption{Mean achievable rates of macrocell UEs and femtocell UEs as functions of the intensity of outside nonsubscribers $\lambda_{\mathrm{out}}$.}
\label{fig:density}
\end{figure}

Figure \ref{fig:density} displays the mean achievable rates of macrocell UEs and femtocell UEs as the outside nonsubscribers intensity $\lambda_\mathrm{out}$ increases.
We observe that the mean achievable rates of macrocell and femtocell UEs drop initially and then tend to be stable.
To interpret this behavior, we note that as the intensity of outside nonsubscribers begins to increase, more subchannels become occupied by MBSs, incurring more macrocell tier interference; however, when the intensity of outside nonsubscribers is sufficiently large, almost all the subchannels are persistently occupied by MBSs with the UEs served by time-sharing, and then the interference saturates thus leading to stable performance for UEs.
In the case when the FAPs are clustered, we observe that the mean achievable rate of femtocell UEs only mildly decreases with the increasing of $\lambda_\mathrm{out}$, suggesting that the performance of femtocell UEs is mainly limited by the interference from the nearby FAPs and has little correlation with the intensity of interfering MBSs.

\begin{figure}
\centering
\includegraphics[width=0.45\textwidth]{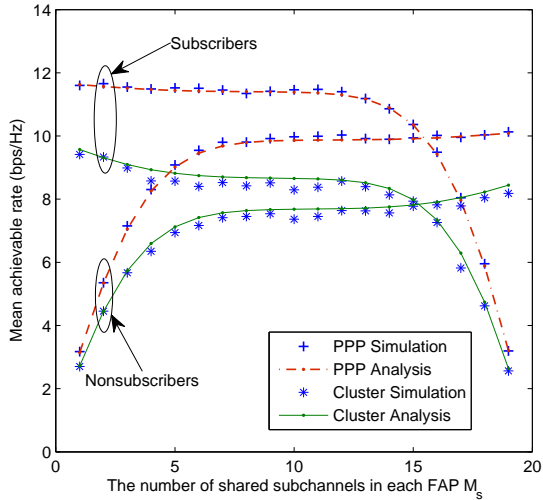}
\caption{Performance of nonsubscribers and subscribers as a function of the number of shared subchannals $M_s$.}
\label{fig:mar1}
\end{figure}

Figure \ref{fig:mar1} displays the mean achievable rates of nonsubscribers and subscribers as functions of the number of shared subchannels,
in which the rate of nonsubscribers is averaged over both outside and inside nonsubscribers.
When few subchannels are to be shared by each FAP (i.e., small $M_s$), the mean achievable rate of nonsubscribers is small while subscribers enjoy a good spectral efficiency.
On the contrary, when most of the subchannels are shared to nonsubscribers, the mean achievable rate of subscribers deteriorates seriously.
Nevertheless, we observe from the figure that there exists a stable compromise at which the rates of both subscribers and nonsubscribers do not drop much from their maxima. For the default configuration in our numerical study, the stable compromise in the PPP case as well as in the clustered case occurs when the value of $M_s$ lies in the range of $[7,12]$, corresponding to a reasonably wide tuning range for system designer when provisioning the resource.

\begin{figure}
  \centering
  \subfigure[$\lambda_s=0.003$ and $\lambda_{\mathrm{in}}=0.027$.]{
    \label{fig:mar23:a} 
    \includegraphics[width=0.45\textwidth]{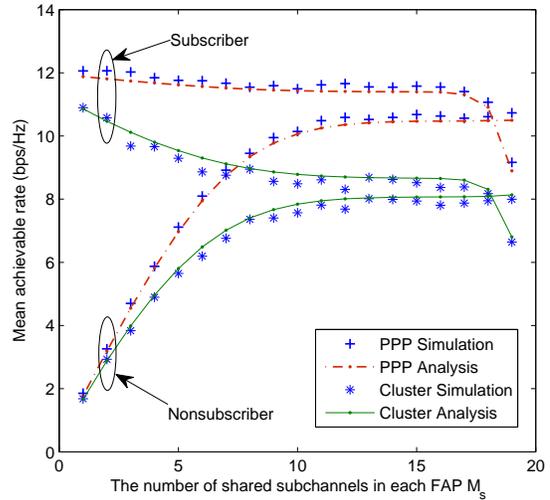}}
  \subfigure[$\lambda_s=0.027$ and $\lambda_{\mathrm{in}}=0.003$.]{
    \label{fig:mar23:b} 
    \includegraphics[width=0.45\textwidth]{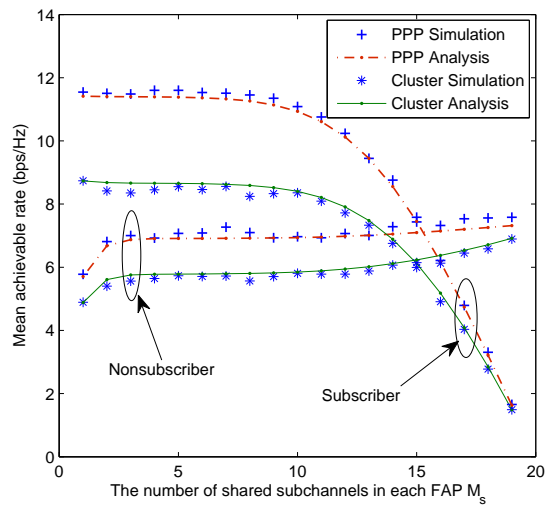}}
  \caption{Performance of nonsubscribers and subscribers with changing proportions of inside nonsubscribers and subscribers.}
  \label{fig:mar23} 
\end{figure}

Figure \ref{fig:mar23} displays the mean achievable rates of nonsubscribers and subscribers with different proportions of inside nonsubscribers and subscribers in femtocells.
For a fair comparison, we fix the sum intensity of the two types of femtocell UEs, as $\lambda_s+\lambda_{\mathrm{in}}=0.03$.
Figure \ref{fig:mar23:a} gives the performance in the case when most of the femtocell UEs are nonsubscribers, while Figure \ref{fig:mar23:b} corresponds to the opponent case.
From the curves, we observe that in order to achieve a good performance for both types of UEs, the number of shared subchannels $M_s$ should be adjusted based on the intensity of inside nonsubscribers.
Moreover, we also find that the tuning range of $M_s$ in the PPP case is almost the same as that in the clustered case; thus illustrating that it is the load of the network rather than the spatial distribution of FAPs that contributes the major impact on the choice of $M_s$.

\section{Conclusions}
\label{sect:conclusions}

In this paper, we explored the application of stochastic geometry in the analysis of hybrid access mechanisms for multi-channel two-tier networks, focusing on the evaluation of the tradeoff between nonsubscribers and subscribers. We characterized several key statistics of UEs and subchannels, and established SINR distribution and mean achievable rate for each type of UEs.

Our analysis revealed the interaction among the various parameters in the network model, and thus shed useful insights into the choice of network parameters and the provisioning of resource in system design. From our numerical study, we observe that although there is an apparent conflict between the interests of nonsubscribers and subscribers, there usually exists a reasonably wide tuning range over which nonsubscribers and subscribers attain a stable compromise at which the rates of both subscribers and nonsubscribers do not drop much from their maxima.
We also found that although the spatial distributions of FAPs are different, the tuning range is almost the same when the intensities of different types of UEs are fixed.
\begin{appendices}

\section{}
\label{appendix:a}

Assume that the typical marcocell UE is located at the origin $o$, and let $r$ be the distance between it and its serving MBS.
Since an outside nonsubscriber always chooses its nearest MBS to access, the cdf of $r$ is obtained by\:
\begin{eqnarray}
\mathbb{P}\{r\leq R\} &=& 1-\mathbb{P}\{\mathrm{no\ MBS\ closer\ than\ }R\}\nonumber\\
&=& 1-e^{-\lambda_m \pi R^2}.
\end{eqnarray}
Then, the pdf of $r$ is $f(r) = e^{-\lambda_m \pi r^2}2\pi\lambda_m r$.
Assuming that the considered UE is at distance $r$ from its serving MBS, $g_\cdot$ is the fading of an interference link, and $R_\cdot$ is the distance between the UE and an interfering access point, the SINR experienced by the UE is $\mathrm{SINR}=\frac{P_mhr^{-\alpha}}{I_m+I_f+\sigma^2}$, where $I_m=\sum_{i\in\Phi_m'\setminus\{b_0\}}P_mg_iR_i^{-\alpha}$ is the interference from the macrocell tier (excluding the serving MBS itself which is denoted by $b_0$) and $I_f=\sum_{j\in\Phi_f'}W P_fg_jR_j^{-\alpha}$ is the interference from the femtocell tier with the wall penetration loss taken into account.

Thus, the cdf of the SINR is given by
\begin{eqnarray}
\!\!\!\!\!\!\!&\!\!\!\!&Z_m(T)=1-\mathbb{P}\{\mathrm{SINR}>T\}\nonumber \\
\!\!\!\!\!\!\!&\!\!\!\!=\!\!\!\!&1-\!\int_0^\infty \!\!\!\!2\pi\lambda_m re^{-\pi\lambda_mr^2}\mathbb{P} \Big\{\frac{P_mhr^{-\alpha}}{I_m+I_f+\sigma^2}>T\Big\}\mathrm{d}r \nonumber \\
\!\!\!\!\!\!\!&\!\!\!\!=\!\!\!\!&1-\!\int_0^\infty \!\!\!\!2\pi\lambda_m re^{-\pi\lambda_mr^2}\mathbb{P} \Big\{h>\frac{Tr^\alpha}{P_m}(I_m+I_f+\sigma^2)\Big\}\mathrm{d}r \nonumber \\
\!\!\!\!\!\!\!&\!\!\!\!=\!\!\!\!&1-\!\int_0^\infty \!\!\!\!2\pi\lambda_m re^{-\pi\lambda_mr^2}\mathbb{E}\Big\{e^{-\frac{\mu Tr^\alpha}{P_m}(I_m+I_f+\sigma^2)}\Big\}\mathrm{d}r \nonumber \\
\!\!\!\!\!\!\!&\!\!\!\!=\!\!\!\!&1-\!\!\int_0^\infty \!\!\!\!\!2\pi\lambda_m re^{-\pi\lambda_mr^2-\frac{\mu Tr^\alpha\sigma^2}{P_m}}\mathcal{L}_{I_m+I_f}\Big(\frac{\mu Tr^\alpha}{P_m}\Big)\mathrm{d}r. \label{equ:z m}
\end{eqnarray}
Now, we evaluate the Laplace transform for the interference conditioning on the fact that the typical UE is served by the nearest MBS $b_0$ which is at the distance $r$.
First, we derive the Laplace transform for the interference of MBSs, denoted as $I_m$.
\begin{eqnarray}
\mathcal{L}_{I_m}(s)&=&E_{\Phi_m'}\Big\{\exp(-s\!\!\!\!\sum_{i\in\Phi_m'\setminus\{b_0\}}\!\!\!\!P_mg_iR_i^{-\alpha})\Big\}\nonumber \\
&=&E_{\Phi_m'}\Big(\!\!\!\!\prod_{i\in\Phi_m'\setminus\{b_0\}}\!\!\!\!E_{g_i}\{\exp(-sg_iR_i^{-\alpha}P_m)\}\Big)\nonumber\\
&=&E_{\Phi_m'}\Big(\!\!\!\!\prod_{i\in\Phi_m'\setminus\{b_0\}}\!\!\!\!\mathcal{L}_g(sR_i^{-\alpha}P_m)\Big), \label{equ:laplace m}
\end{eqnarray}

The probability generating functional of PPP $\Phi$ in the region $D$ with intensity $\lambda$, denoted by $G_p(v)=\mathbb{E}\{\prod_{x\in\Phi}v(x)\}$, is given by \cite{stoyan1995stochastic} as follows
\begin{equation}
G_p(v) = \exp\Big(-\lambda\int_{D}(1-v(x))\mathrm{d}x\Big).
\end{equation}

Let $C(o,r)$ be the circle centered at the origin $o$ with radius $r$.
As there is no MBS in the circle $C(o,r)$, we have $\Phi_m'(C(o,r))=\emptyset$.
This implies that the interfering MBSs are distributed as PPP on the space $R^2$ exclusive of the region $C(o,r)$.
Let $D=R^2\setminus C(o,r)$ and $v(x)=\mathcal{L}_g(s|x|^{-\alpha}P_m)$, by applying the probability generating functional of PPP we get
\begin{eqnarray}
\!\!\!\!\!\!\!\!\!&\!\!\!\!\!\!\!\!&\mathcal{L}_{I_m}(s)=\exp\Big(-2\pi\lambda_m'\int_r^\infty(1-\mathcal{L}_g(sx^{-\alpha}P_m))x\mathrm{d}x\Big) \nonumber\\
&\!\!\!\!\!\!\!\!&=\exp\Big(-2\pi\lambda_m'\underbrace{\int_0^\infty\!\!\int_r^\infty\!\!(1-e^{-sx^{-\alpha}P_mg})f(g)x\mathrm{d}x\mathrm{d}g}_{(A)} \Big). \nonumber\\
&\!\!\!\!\!\!\!\!&\label{eqn:laplace I_m temp}
\end{eqnarray}
where the last equation follows from the exchange of the integral order.

Let $y=sP_mgx^{-\alpha}$ and integrate by parts, from the properties of Gamma function we obtain
\begin{eqnarray}
(A)=-\frac{r^2}{2}+\frac{1}{\alpha}(sP_m)^{\frac{2}{\alpha}}\mathbb{E}_g\Big\{\nonumber\\
g^{\frac{2}{\alpha}}\Big(\Gamma(-\frac{2}{\alpha},sP_mgr^{-\alpha})-\Gamma(-\frac{2}{\alpha})\Big)\Big\}.
\end{eqnarray}

The evaluation of the Laplace transform for $I_f$ is almost the same except that the interfering FAPs are distributed on the whole space $R^2$. Similar to (\ref{eqn:laplace I_m temp}), we get
\begin{eqnarray}
\mathcal{L}_{I_f}(s)&\!\!\!\!=\!\!\!\!&\exp\Big(-2\pi\lambda_f'\int_0^\infty\!\!(1-\mathcal{L}_g(sx^{-\alpha}WP_f))x\mathrm{d}x\Big) \nonumber\\
&\!\!\!\!=\!\!\!\!&\frac{1}{2}(sWP_f)^{\frac{2}{\alpha}}\Gamma\Big(1-\frac{2}{\alpha}\Big)\mathbb{E}_g\Big\{g^{\frac{2}{\alpha}}\Big\}.\label{eqn:laplace PPP}
\end{eqnarray}
Substituting $\mathcal{L}_{I_m}(s)$ and $\mathcal{L}_{I_f}(s)$ into $Z_m(T)$ hence leads to (\ref{equ:OUT_SINR_DIS}).

Now we evaluate the mean achievable rate. Since for a positive random variable $X$, $\mathbb{E}\{X\}=\int_{t>0}\mathbb{P}\{X>t\}\mathrm{d}t$, we have
\begin{eqnarray}
\tau_m&=&\mathbb{E}\{\ln(1+\mathrm{SINR})\}\nonumber\\
&=&\int_0^\infty\mathbb{P}\Big\{\ln(1+\mathrm{SINR})>t\Big\}\mathrm{d}t\nonumber\\
&=&\int_0^\infty\mathbb{P}\Big\{\mathrm{SINR}>e^t-1\Big\}\mathrm{d}t\nonumber\\
&=&\int_0^\infty(1-Z_m(e^t-1))\mathrm{d}t. \label{equ:tau_m_tmp}
\end{eqnarray}
Plugging $Z_m(T)$ into (\ref{equ:tau_m_tmp}), we arrive at (\ref{equ:OUT_RATE}) and thus establish Theorem \ref{thm:sinr macrocell}.

\section{}
\label{appendix:c}

Assume that the typical UE is located at the origin $o$.
Let $r$ be the distance between a femtocell UE and its serving FAP.
Because the femtocell UEs are uniformly distributed in the circular coverage area of radius $R_f$ of each FAP, the pdf of $r$ is given by $f(r) = {2 r}/{R_f^2}$.

Denote by $I_m=\sum_{i\in\Phi_m'}WP_mg_iR_i^{-\alpha}$ and $I_f=\sum_{j\in\Phi_f'\setminus\{b_0\}}W^2 P_fg_jR_j^{-\alpha}$ the interference strengths from MBSs and FAPs respectively.
Similar to the derivation of (\ref{equ:z m}), we have
\begin{equation}
Z_f(T) = 1-\int_0^{R_f} \frac{2r}{R_f^2} e^{-\frac{\mu Tr^\alpha\sigma^2}{P_f}}\mathcal{L}_{I_m+I_f}\Big(\frac{\mu Tr^\alpha}{P_f}\Big)\mathrm{d}r. \label{equ:Zf}
\end{equation}
Since $\Phi_m'$ is a homogeneous PPP with intensity $\lambda_m'$, we obtain the Laplace transform for $I_m$ similar as the derivation of (\ref{eqn:laplace PPP})
\begin{equation}
\!\!\!\!\mathbb{E}\big\{e^{-sI_m}\big\}=\exp\Big(-\pi\lambda_m'(sWP_m)^{\frac{2}{\alpha}}\Gamma\Big(1-\frac{2}{\alpha}\Big)\mathbb{E}_g(g^{\frac{2}{\alpha}})\Big).
\end{equation}
The FAPs are distributed as a homogeneous PPP; however, the serving FAP is not included when calculating the interference.
By the Slivnyak-Mecke Theorem, the reduced Palm distribution of the Poisson p.p. is equal to its original distribution.
Thus, the Laplace transform for $I_f$ can still be obtained similar as the derivation of (\ref{eqn:laplace PPP})
\begin{equation}
\!\!\!\!\mathbb{E}\big\{e^{-sI_f}\big\}=\exp\Big(-\pi\lambda_f'(sW^2P_f)^{\frac{2}{\alpha}}\Gamma\Big(1-\frac{2}{\alpha}\Big)\mathbb{E}_g(g^{\frac{2}{\alpha}})\Big).
\end{equation}

In the above, it is noteworthy that the interference from an FAP penetrates two walls thus the loss becoming $W^2$ instead of $W$.
Substituting the Laplace transform for $I_m$ and $I_f$ into (\ref{equ:Zf}) with $v=r^2$, we get the SINR distribution,
and similar to (\ref{equ:tau_m_tmp}), we get the mean achievable rate.

\section{}
\label{appendix:e}

The derivation is exactly the same as Appendix \ref{appendix:a} till the equation (\ref{equ:z m}).
The essential distinction of the derivation lies in that the Laplace transform for the interference is different from that in the PPP case.
Let $I_m=\sum_{i\in\Phi_m'\setminus\{b_0\}}P_mg_iR_i^{-\alpha}$ and $I_f=\sum_{j\in\Phi_f'}W P_fg_jR_j^{-\alpha}$.
Since there is no MBS in the disk $C(o,r)$, we have $\Phi_m'(C(o,r))=\emptyset$.
Referring to the previous derivation in the PPP case, we obtain the Laplace transform for $I_m$ as follows
\begin{equation}
\mathbb{E}\big\{e^{-sI_m}\big\}=\exp\Big(-\pi\lambda_m'r^2\varphi\Big(\frac{sP_m}{\mu r^\alpha},\alpha\Big)\Big)
\end{equation}
The FAPs are distributed as a Neyman-Scott cluster process and the generating functional $G(v)=\mathbb{E}(\prod_{x\in\Phi}v(x))$ is given by \cite[Page 157]{stoyan1995stochastic}
\begin{eqnarray}
G(v)=\exp\Big(-\lambda_p\int_{R^2}\Big(1-\nonumber\\
\exp\Big(-\lambda_c\int_{C(o,R_c)}(1-v(x+y))\mathrm{d}y\Big)\Big)\mathrm{d}x\Big) \label{equ:generating functional}
\end{eqnarray}
Similar to the derivation of (\ref{equ:laplace m}), we get the Laplace transform for $I_f$
\begin{eqnarray}
\mathcal{L}_{I_f}(s)&\!\!\!\!=\!\!\!\!&\mathbb{E}\Big\{\prod_{j\in\Phi_f'}\mathcal{L}_g(sR_j^{-\alpha}WP_f)\Big\}\nonumber\\
&\!\!\!\!=\!\!\!\!&\mathbb{E}\Big\{\prod_{j\in\Phi_f'}\frac{\mu}{\mu+sR_j^{-\alpha}WP_f} \Big\} \label{eqn:laplace I_f cluster}
\end{eqnarray}
Let $v(x)=\frac{\mu}{\mu+sP_fW|x|^{-\alpha}}$ and plugging into the generating functional of Neyman-Scott cluster process (\ref{equ:generating functional}), we get the Laplace transform for $I_f$.
Having derived the Laplace transform for $I_m$ and $I_f$, similar to the derivations in Appendix \ref{appendix:a}, we obtain the results in Theorem \ref{thm:sinr macrocell_cluster}.

\section{}
\label{appendix:f}

Different from the above proofs, we assume that the serving FAP rather than the typical UE is located at the origin.
The typical UE is distributed in the circle centered at the origin with radius $R_f$.
Let $I_m=\sum_{i\in\Phi_m'}WP_mg_iR_i^{-\alpha}$ and $I_f=\sum_{j\in\Phi_f'\setminus\{b_0\}}W^2 P_fg_jR_j^{-\alpha}$.

First, we evaluate the Laplace transform for the interference conditioned on the fact that the typical UE is located at distance $r$ from the serving FAP located at the origin.
Without loss of generality, we assume that the typical UE is located at $z=(r,0)$.
Since $\Phi_m'$ is a homogeneous PPP with intensity $\lambda_m'$, we obtain the Laplace transform for $I_m$ similar as the derivation of (\ref{eqn:laplace PPP})
\begin{equation}
\!\!\!\!\!\!\mathbb{E}\big\{e^{-sI_m}\big\}=\exp\Big(-\pi\lambda_m'\Big(\frac{sWP_m}{\mu}\Big)^{\frac{2}{\alpha}}\Gamma(1+\frac{2}{\alpha})\Gamma(1-\frac{2}{\alpha})\Big)
\end{equation}
The FAPs are distributed as a Neyman-Scott cluster process.
Since the serving FAP, located at the origin, is not included when calculating the interference,
we should consider the reduced Palm distribution of the cluster process when evaluating the Laplace transform for $I_f$.
Let $G_o^!(v)=\mathbb{E}_o^!(\prod_{x\in\Phi}v(x))$ denotes the generating functional of the reduced Palm distribution of the cluster process.
The notation $\mathbb{E}_o^!(\cdot)$ denotes the conditional expectation for the point process given that there is a point of the process at the origin but without including the point.
The conditional generating functional $G_o^!(v)$ is given by the Lemma 1 in \cite{ganti2009interference} as follows
\begin{eqnarray}
G_o^!(v)=\frac{1}{\pi R_c^2}G(v)\int_{C(o,R_c)}\nonumber\\
\exp\Big(-\lambda_c\int_{C(o,R_c)}(1-v(x-y))\mathrm{d}x\Big)\mathrm{d}y\label{equ:conditional generating functional}
\end{eqnarray}
where $G(v)$ is the generating functional of Neyman-Scott cluster process and is given by (\ref{equ:generating functional}).

Referring to (\ref{eqn:laplace I_f cluster}), let $v(x)=\frac{\mu}{\mu+sP_fW|x-z|^{-\alpha}}$. Plugging $v(x)$ into the conditional generating functional of Neyman-Scott cluster process (\ref{equ:conditional generating functional}), we get the Laplace transform for $I_f$.
Having derived the Laplace transform for $I_m$ and $I_f$, similar to the derivation in Appendix \ref{appendix:c},
we obtain the results in Theorem \ref{thm:sinr femtocell_cluster}.

\end{appendices}
\section*{Acknowledgement}

The authors wish to thank the anonymous reviewers for their valuable comments on this work.

\bibliographystyle{IEEEtran}
\bibliography{123}

\balance
\clearpage
\end{document}